\documentclass[usenatbib]{mn2e}
\usepackage{graphicx,natbib}
\usepackage{amsmath}

\newcommand{\apj}{ApJ}
\newcommand{\apjl}{ApJ}

\newcommand{\mnras}{MNRAS}

\newcommand{\aap}{A\&A}

\newcommand{\msun}{M_{\odot}}\newcommand{\tmsun}{\mbox{$\msun$}}

\newcommand{\myvec}[1]{{\boldsymbol #1}}
\newcommand\simless{\mathbin{\lower 3pt\hbox
   {$\rlap{\raise 5pt\hbox{$\char'074$}}\mathchar"7218$}}}
\newcommand\simgreat{\mathbin{\lower 3pt\hbox
   {$\rlap{\raise 5pt\hbox{$\char'076$}}\mathchar"7218$}}}

\title[Young stellar disc(s) in the Galactic Centre]
      {On the number of young stellar discs in the Galactic Centre}

\author[U. L\"ockmann and H. Baumgardt]
{
  U. L\"ockmann\thanks{E-mail: uloeck@astro.uni-bonn.de (UL); holger@astro.uni-bonn.de (HB)}
  and H. Baumgardt\footnotemark[1]\\ 
  Argelander Institute for Astronomy, University of Bonn, Auf dem H\"ugel 71, 53121 Bonn,
  Germany\\
}

\begin{document}

\date{Accepted 2009 January 8.  Received 2009 January 6; in original form 2008 June 16}

\pagerange{\pageref{firstpage}--\pageref{lastpage}} \pubyear{2008}

\maketitle

\label{firstpage}

\begin{abstract}
Observations of the Galactic Centre show evidence of disc-like structures of very young stars orbiting the central super-massive black hole within a distance of a few 0.1 pc.
While it is widely accepted that about half of the stars form a relatively flat disc rotating clockwise on the sky,
there is a substantial ongoing debate on whether there is a second, counter-clockwise disc of stars.

By means of $N$-body simulations using our \textsc{bhint} code, we show that two highly inclined stellar discs with the observed properties cannot be recognised as two flat circular discs after 5\,Myr of mutual interaction.
Instead,
our calculations predict a significant warping of the two discs, which we show to be apparent among the structures observed in the Galactic Centre.
While the high eccentricities of the observed counter-clockwise orbits suggest an eccentric origin of this system, we show the eccentricity distribution in the inner part of the more massive clockwise disc to be perfectly consistent with an initially circular disc in which stellar eccentricities increase due to both non-resonant and resonant relaxation.

We conclude that the relevant question to ask is therefore not whether there \emph{are} two discs of young stars, but whether there \emph{were} two such discs to begin with.

\end{abstract}


\section{Introduction}

Observations of the Galactic Centre (GC) revealed one or two discs of 5-6\,Myr old stars orbiting the central super-massive black hole (SMBH)
at a distance of $\sim$0.1\,pc \citep[and references therein]{pau06}.
The presence of such very young stars in the close vicinity of the SMBH is puzzling, and a number of processes have been proposed to resolve this \emph{Paradox of Youth}:
\citet{ger01} suggested that a disc of stars may have formed by tidal disruption of an infalling cluster of young stars, which would require a large mass or a central intermediate-mass black hole to survive the strong tidal forces from the SMBH \citep{mp03}. However, \citet{lb03} showed that this cannot explain the high proximity of the stars, and propose in-situ formation by fragmentation of a massive accretion disc as an alternative scenario \citep[see also][]{nc05}.
\citet{mhmw08} and \cite{br08} have shown the infall of a molecular cloud towards the GC to be effective in creating a disc of stars with distances from Sgr A$^*$ consistent with the observed young massive stars.

While the disc of stars orbiting clockwise around the SMBH in the centre of our Galaxy is very distinct, the disc structure of the counter-clockwise system is still under discussion \citep{lu06,lu08}. However,
\citet{caa08} have shown that a single cold disc in the absence of a perturbing potential cannot explain the observed large inclinations and eccentricities.
The required perturbing potential may result from a massive circum-nuclear disc \citep{ght94}, an intermediate-mass black hole and/or star cluster \citep[probably IRS 13;][]{mpsr04}, or a second disc \citep{pau06}.
As there is evidence for a second, counter-rotating disc, and it has been shown that disc dissolution is effective enough by mutual interaction between two discs \citep{ndcg06}, this is currently the most promising solution suggested.
We therefore test the hypothesis that two discs of stars formed almost simultaneously a few Myr ago against the observations \citep[see also review by][]{a05}.

\section{Numerical method and models}

To test the mutual interaction of two stellar discs orbiting the central SMBH, we performed direct {\itshape N}-body simulations using our {\small BHINT} code \citep{lb08}, which has been developed specifically to calculate the dynamics of stars orbiting around a SMBH. It includes post-Newtonian approximation of the effects of general relativity up to order 2.5. However, test runs show that these corrections do not have a major impact on the results discussed in this work, as the stars hardly reach eccentricities  $e>0.9$.
The period of relativistic precession about a $3.5 \times 10^6\,\msun$ SMBH is \citep{e16}
\begin{equation}
  P_{\rm prec,GR} = 2 \times 10^5 P_{\rm orb} \left(\frac{a}{0.1 \rm pc}\right)\left(1-e^2\right), \label{eq:prec-gr}
\end{equation}
where $P_{\rm orb}$, $a$ and $e$ are the orbital period, semi-major axis, and eccentricity, respectively,
while the period of precession due to a spherical cusp as observed by \citet{sea+07} is given by \citep{ips05}
\begin{equation}
  P_{\rm prec,cusp} = 73 P_{\rm orb} \left(\frac{a}{0.1 \rm pc}\right)^{-1.8}\left(1-e^2\right)^{-0.5}. \label{eq:prec-cusp}
\end{equation}
Dividing the two yields
\begin{equation}
  \frac{P_{\rm prec,GR}}{P_{\rm prec,cusp}} = 2.7 \times 10^3 \left(\frac{a}{0.1 \rm pc}\right)^{2.8}\left(1-e^2\right)^{1.5},
\end{equation}
which for $a>0.03$\,pc (around the minimum achieved for the disc stars in our simulations) and $e<0.9$ is larger than $\approx 8$, showing that precession is dominated by the cusp effect in the scope of this Letter (here we will not care about the actual distribution of eccentricities $e>0.9$), which we also find in our simulations.
Therefore we waive post-Newtonian treatment to keep computing times low.

Mass ratio and distributions of the stars' orbital parameters are chosen such that the final state is consistent with the observed values. We start with two cold stellar discs with total masses of 10,000 and 5,000\,\tmsun, respectively, inclined at an angle of 130$^{\circ}$ (run $A$).
Similar values have been shown by \citet{ndcg06} to be sufficient in leading to an extent of disc dissolution comparable to the observations.
The discs have a finite radial extent of $0.05-0.5$\,pc and $0.07-0.5$\,pc, respectively,
a surface density $\Sigma (R) \propto R^{-2.5}$, and a thickness (standard deviation) of 1.44$^{\circ}$. The stellar orbits have a mean eccentricity of 0.03.
The underlying initial mass function (IMF) has a slope of $\gamma = 1.35$, as derived by \citet{pau06}.
Since the disc mass for such a flat IMF is not very sensitive to the lower stellar mass limit, we use a mass range of $1-120\,\msun$.

The presence of a stellar cusp as it is observed in the GC \citep[e.g.,][]{sea+07} leads to pericentre shift of eccentric orbits, while the orbital plane (and thus any disc warping) is conserved. As \citet{ips05} pointed out, pericentre shift prevents high eccentricities achieved by the Kozai effect.
To correctly model the disc evolution,
we accounted for the influence of a background cusp by adding 14,000 stellar mass black holes (SBHs) of 15\,\tmsun\ into the inner 0.22\,pc, consistent with the density profile derived by \citet{sea+07}.
The use of 15\,\tmsun\ black holes is consistent with \citet{fak06} finding that this region is mass-dominated by the most massive remnants having migrated into the centre by dynamical friction during the lifetime of the Galaxy.

Outside 0.22\,pc, the eccentricities achieved are only moderate, thus the effects of pericentre shift are expected to be low. To verify this, we calculated models including an analytic potential or a cusp of 40\,\tmsun\ SBHs outside 0.22\,pc, which did not yield significantly different results.

\section{Results}

\subsection{Disc warping}
The precession frequency of a star orbiting a SMBH of mass $M_{\rm SMBH}$ at radius $R$ at an inclination $\beta$ relative to a narrow disc of mass $M_{\rm disc}$ at a radius $R_{\rm disc}$ can be approximated as \citep{nay05}
\begin{equation}
\omega_p = -\frac{3}{4}\frac{M_{\rm disc}}{M_{\rm SMBH}}\cos{\beta}\sqrt{\frac{GM_{\rm SMBH}}{R^3}}\frac{R^3R_{\rm disc}^2}{\left(R^2+R_{\rm disc}^2\right)^{5/2}}, \label{eq:warp}
\end{equation}
which for our model roughly translates to a precession period of $P_{\rm prec,disc} \sim 5\,{\rm Myr} \left(R/0.1\,{\rm pc}\right)^{-3/2}$ for stars in the less massive disc (about the massive disc), and twice as long for stars in the massive disc.
Note, however, that this is only a very rough approximation and is only valid if the disc is not radially extended and $R \ll R_{\rm disc}$.
For the evolution of two massive discs, the situation is further complicated by precession of the reference disc, precession of stars about their disc of origin and the resulting increase of $\left|\cos{\beta}\right|$, which makes detailed predictions impossible.

As can be seen, the precession frequency is proportional to the other disc's mass, but also depends on the distance of a star from the central SMBH.
Since both discs have a finite extent, stars at different central distances precess with different frequencies,
thus warping the originally flat disc.

\begin{figure}
  \begin{center}
    \includegraphics[width=8.3cm]{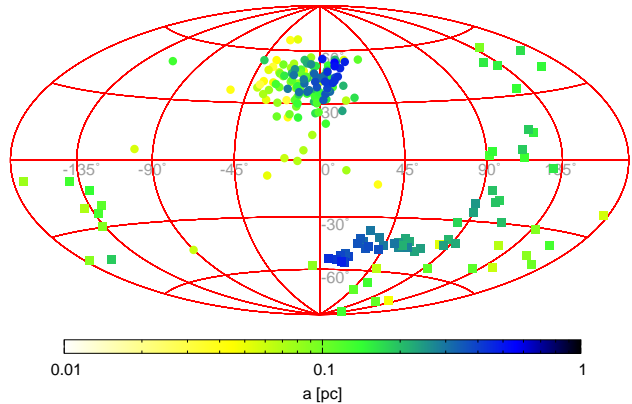}
  \end{center}
  \caption{Orientation of stellar orbits as a function of semi-major axis $a$ after 5\,Myr of simulation.
    The compact shape of filled circles indicates moderate warping of the massive disc,
    while the elongated structure of squares shows a massive warping until disc dissolution towards the inner edge of the less massive disc.
    \label{fig:warp-sim}}
\end{figure}

Fig.\ \ref{fig:warp-sim} shows the Aitoff projection of the normal vectors to the stars' orbital planes after 5\,Myr of simulation.
The compact shape of the massive disc's normal vectors indicates the moderate warping of this disc. This also holds for the outer part of the less massive disc, while the elongated structure at smaller radii shows a massive warping until disc dissolution towards the inner edge of this disc.

The observed GC discs show moderate thickness \citep{pau06}. To illustrate the present warping, we fit discs to subsets of the stars at different radii.
Having grouped the stars into a clockwise and a counter-clockwise system according to \citet{pau06}, we sort each group by projected radius, and fit a disc to the velocity vectors of any 6 subsequent members of the two sorted lists.

\citet{lb03} fitted discs to the GC stars using a $\chi^2$ test based on the observed velocity vectors and three-dimensional error bars.
For better comparison between observations and simulations, where we have no error bars, we fit discs by minimising the root mean square angle $\alpha$ between the stars' velocity vectors $\myvec{v_i}$ and the fitted plane,
\begin{equation}
  \alpha = \sqrt{\frac{1}{N}\sum_{i=1}^N \arcsin^2 \myvec{n}\myvec{v_i}},
\end{equation}
where $\myvec{n}$ is a normal vector on the fitted plane. $\alpha$ can be regarded as the thickness of the disc.
To account for observational uncertainties, we draw the three components of the stars' velocity vectors from a Gaussian using the respective observed velocity component and observational error as the expectation value and standard deviation, respectively.
The actual orientation of an observed disc is thus retrieved as the mean of the resulting normal vectors from 2000 such realisations.

\begin{figure}
  \begin{center}
    \includegraphics[width=8.3cm]{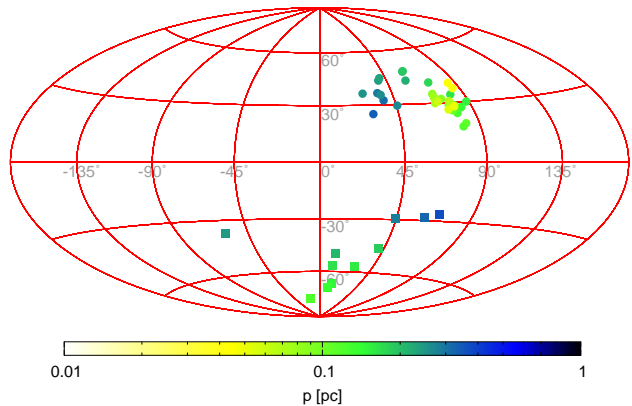}
  \end{center}
  \caption{Orientation of stellar orbits in the GC as a function of projected central distance $p$. Symbols as in Fig.\ \ref{fig:warp-sim}. We find a compact spiral structure for the massive (clockwise) disc, and an elongated structure depicting the dissolving counter-clockwise disc.
    \label{fig:warp-obs}}
\end{figure}

Fig.\ \ref{fig:warp-obs} shows the fitted normal vectors for the observed clockwise and counter-clockwise system. While the massive clockwise disc shows a compact spiral structure, demonstrating the overall flatness of this disc, the elongated shape of the less massive (counter-clockwise) system hints at the fact that it loses its disc shape towards the centre, as the orientation is a strong function of central distance.

The warp direction in the observations appears to be clockwise for decreasing central distance, just opposite to the results from our simulation as depicted in Fig.\ \ref{fig:warp-sim}. This fact suggests that the two discs started with an \emph{initial} inclination of less than 90$^{\circ}$.
We therefore integrated a second model, with an initial inclination of 88$^{\circ}$ (run $B$).

\begin{figure}
  \begin{center}
    \includegraphics[width=8.3cm]{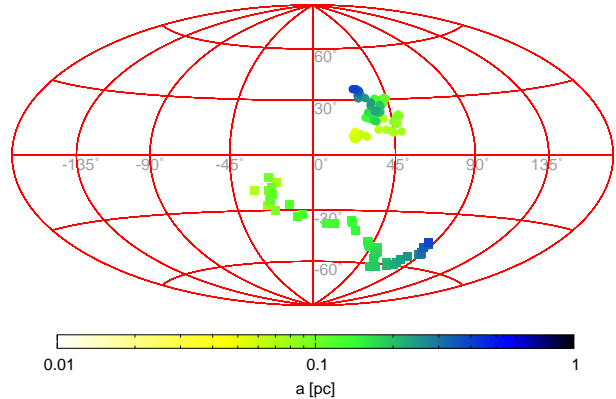}
  \end{center}
  \caption{Results from run $B$ with an initial inclination of 88$^{\circ}$. This time, we plot fitted normal vectors obtained in the same way as for the observations. The concordance of direction and shape of the elongated structure for the less massive disc is obvious. We also find a compact spiral for the clockwise disc similar to the observations, although with slightly different orientation.
    \label{fig:warp-sim2}}
\end{figure}

For better comparison with the observations, we now sort the massive stars of either disc by semi-major axis, and again fit discs to the velocity vectors of now any 12 subsequent stars (since we have more stars available than in the observations). The results are shown in Fig.\ \ref{fig:warp-sim2}. Again, we find an elongated structure for the counter-clockwise system, this time consistent with the orientation of the observed orbits as in Fig.\  \ref{fig:warp-obs}.
The compact image of the clockwise disc exhibits a spiral structure similar to the observed disc, although with slightly different orientation.
The diffusion towards the inner edge of this disc is due to fast precession of the corresponding stellar orbits about their disc of origin.

\begin{figure}
  \begin{center}
    \includegraphics[width=8.3cm]{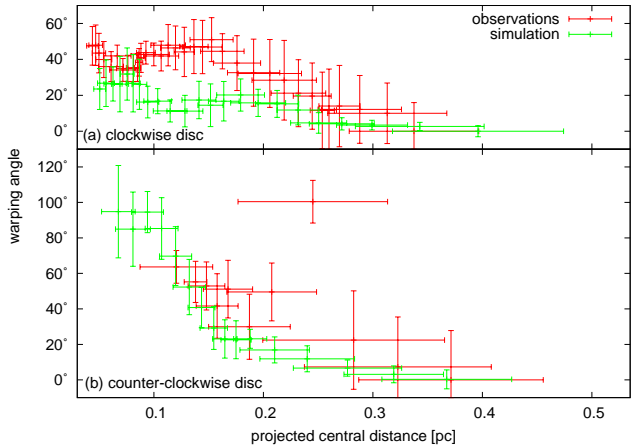}
  \end{center}
  \caption{Disc warping angle as a function of central distance. We plot the angle between the plane fitted to any disc's subset of stars and the outermost fitted plane for the respective disc. The upper panel shows the results for the clockwise disc from both observations (red) and simulations (run $B$; green); the lower panel shows the respective results for the counter-clockwise systems.
  Horizontal and vertical error-bars indicate standard deviation of central distance and 1$\sigma$-thickness of the fitted planes, respectively.
  It is evident that the angle does not randomly scatter, but increases with decreasing central distance, thus showing the disc warping.
    \label{fig:warp-meas}}
\end{figure}

To quantify the amount of warping of the stellar discs, we can measure the angle between the planes fitted to two subsets of stars at different radii within one disc. As we do not have the original orientation of the observed discs, we take the orientation of the outermost fitted plane as reference, since according to Eq.\ \ref{eq:warp} and our simulation results, the outermost stars show the least warping.
Fig.\ \ref{fig:warp-meas} compares the thus calculated warping angles of the observations and our run $B$.
It can be seen that the warping is higher for the less massive disc, especially towards the inner edge. This also explains the lack of observed counter-clockwise disc stars towards the inner edge.
Also, \citet{lu08} do not find any counter-clockwise disc structure in their improved measurements within 0.14\,pc from Sgr A$^*$, which due to the strong warping effect would not be expected, as our simulations show.

Averaging over five realisations of our model, we retrieve maximum warping angles of $28^{\circ} \pm 4^{\circ}$ for the clockwise and $108^{\circ} \pm 19^{\circ}$ for the counter-clockwise disc. While the warping angles of the counter-clockwise disc fit the observations quite well,
the observed warping of the clockwise disc is higher than in our simulations, which may be due to a larger mass in the other disc, or any other perturbing potential like star clusters or the circum-nuclear disc.

Similar to what we have seen in Fig. \ref{fig:warp-sim2}, the dips towards the inner edge of the massive clockwise discs can be ascribed to fast precession of the corresponding orbits about their disc of origin.

\subsection{Relaxation in the Galactic Centre}
\label{sec:err}

As we have shown in the previous section, the precession caused by the stellar discs is the main driver for the change of the stars' orbital inclination. However, it does not change orbital eccentricity or semi-major axis. These properties are changed by relaxation processes among the stars in the GC.

The relaxation time of a stellar system is known as the time it takes for the stars to change their energy (or angular momentum) by a factor of order unity and can be estimated as \citep[e.g.,][Eq. (2-62)]{s87}
\begin{equation}
  t_{\rm rel} = \frac{0.065 \sigma^3}{G^2 M_* \rho \ln \Lambda},
\end{equation}
where $\sigma$ is the local 3D velocity dispersion, $\rho$ is the stellar mass density, and $\ln \Lambda$ is the Coulomb logarithm. Using the orbital circular velocity around a $3.5 \times 10^6\,\msun$ SMBH for $\sigma$, the stellar mass profile by \citet{sea+07} at central distance $r<0.22$\,pc, and $\ln \Lambda \sim 10$, this gives
\begin{equation}
  t_{\rm rel} \sim 190\,{\rm Myr} \left(\frac{r}{0.1\,{\rm pc}}\right)^{-0.3}\left(\frac{M_*}{15\,\msun}\right)^{-1}.
\end{equation}
The mean squared change of energy $E$ and angular momentum $L$
of a star after a time $t$ is described by
\begin{equation}
  t/t_{\rm rel} = \left<\left(\Delta E/E_0\right)^2\right>
                = \left<\left(\Delta L/L_{\rm c,0}\right)^2\right>
\end{equation}
\citep[eq. (4)]{rt96}, where $E_0$ is the initial energy and $L_{\rm c,0}$ the angular momentum of a corresponding circular orbit.
This way, we can expect the stellar orbits in the GC to be fully randomised only after $\sim 200$\,Myr, and only some stars on moderate eccentricities after 5\,Myr.
However, relaxation is enhanced within the stellar discs:
While the stellar discs do not contribute much to the average density of the cusp, they significantly increase the \emph{local} density in the vicinity of the disc stars. In our models, a sphere of radius $r_s \ll 0.1$\,pc centred on a star in the massive disc at 0.1\,pc from the SMBH contains $3\times 10^7 \left(r_s/{\rm pc}\right)^3 \msun$ of cusp stars and $2.6 \times 10^5 \left(r_s/{\rm pc}\right)^2 \msun$ of disc stars, hence for $r_s \sim 0.01$\,pc the disc stars double the local density and thus the amount of relaxation.
The small relative velocities in the initially cold disc are expected to further enhance this effect.

\begin{figure}
  \begin{center}
    \includegraphics[width=8.3cm]{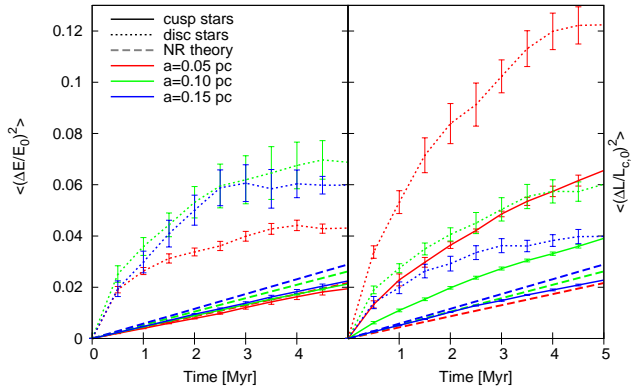}
  \end{center}
  \caption{Time evolution of mean squared energy and angular momentum change for model $B$, averaged over five different realisations with standard deviation shown as error bars. The normalisation constants $E_0$ and $L_{\rm c,0}$ are the local energy and angular momentum of a circular orbit, respectively.
  It can be seen that relaxation in the discs (dotted lines) is well enhanced, most likely due to local over-density and small relative velocities.
  Energy relaxation (left) of the cusp stars follows the theoretical prediction of non-resonant relaxation very well, showing only weak dependence on central distance. The apparent radial dependence of energy relaxation in the discs may be explained by faster decrease of local over-density as a consequence of stronger warping at the inner edge of the disc (dotted red line).
  On the other hand, relaxation of angular momentum (right) shows a strong radial dependence and appears to be much faster than energy relaxation towards the centre.
  We thus conclude that an additional component of angular momentum relaxation must be present, i.e.\ resonant relaxation.
    \label{fig:rel}}
\end{figure}

Fig.\ \ref{fig:rel} shows the evolution of mean squared change of energy and angular momentum 
for the stars of run $B$ for different initial semi-major axes. 
While the cusp stars follow the theoretical prediction of energy relaxation very well, it is clearly seen that relaxation is faster for the disc stars, especially so in the beginning, where the local density is at its maximum.
Furthermore, angular momentum relaxation appears to be much faster than predicted by non-resonant relaxation and shows a strong dependence on central distance, suggesting that resonant relaxation \citep[RR;][]{rt96} plays an important role.

The RR time-scale is given by \citet[eq.\ (9)]{rt96} as
\begin{equation}
  t_{\rm RR} \sim P_{\rm orb} \frac{M_{\rm SMBH}}{M_*}
             \sim 370\,{\rm Myr} \left(\frac{r}{0.1\,{\rm pc}}\right)^{1.5}\left(\frac{M_*}{15\,\msun}\right)^{-1},
\end{equation}
hence, while the contribution of RR at 0.1\,pc is only half as much as for non-resonant relaxation (NR), it is expected to dominate relaxation further in.
Replacing the stellar mass of $M_*=15\,\msun$ by the Paumard mass function used for the discs would increase RR efficiency by a factor of 3, since $M_*$ needs to be replaced by $\int m^2 {\rm d}N(m)/\int m {\rm d}N(m) \sim 50\,\msun$. The alignment of stellar orbits in two discs may further accelerate RR, as the sum of torques exerted causing a change in angular momentum is no longer random.
Replacing the stellar cusp by an analytic potential, we find that angular momentum relaxation as shown in Fig.\ \ref{fig:rel} is decreased by $\sim 20\%$, but still significantly higher than that of the cusp stars in the previous simulation, confirming that while the cusp stars do contribute, relaxation among disc stars is dominant.

Note that RR is only fully effective as long as the orbits are almost stationary over some time $t$ such that $P_{\rm orb} \ll t \ll P_{\rm prec}$, which according to Eq.\ \ref{eq:prec-cusp} is only fulfilled in the inner part of the discs; however, in the outer parts of the disc the theoretical RR time-scale is much longer than the NR one, so that resonant effects cannot be expected to significantly contribute to stellar relaxation.

\begin{figure}
  \begin{center}
    \includegraphics[width=8.3cm]{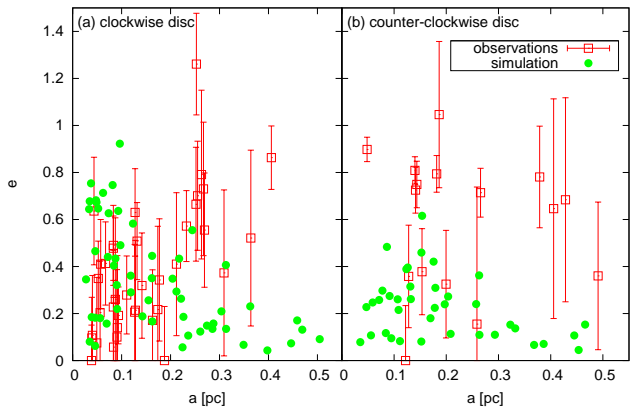}
  \end{center}
  \caption{Eccentricity of massive disc stars as a function of (projected) central distance.
    The filled circles show the results of run $B$ after a simulated time of 5\,Myr,
    the open squares depict the observations as given by \citet{pau06}.
    Observational errors are denoted as 1$\sigma$ error bars.
    While the eccentricities observed in the clockwise disc (left panel)
    within a distance of $< 0.3$\,pc from the SMBH can be well explained by our model,
    the orbits of stars towards the outer edge of the population
    (having an admittedly high uncertainty) may 
    need a different
    explanation, such as a more complex infall geometry, or a massive perturber.
    \label{fig:ecc}}
\end{figure}

In the following, we analyse the distribution of the disc stars' orbital parameters resulting from relaxation as described above.
Fig.\ \ref{fig:ecc} shows the eccentricities of massive disc stars ($>17\,\msun$) after 5\,Myr of simulation as a function of semi-major axis.
It can be seen that a considerable fraction of stars from both discs achieve eccentricities up to $e=0.9$. 
This shows that the observation of young stars on highly eccentric orbits is as such not a valid argument against a circular disc origin of these stars.
However, this is only true for
the inner part of the disc, while most stars beyond 0.3\,pc from the SMBH do not show eccentricities larger than $e=0.2$ in our runs.
Since this might be explained by the absence of cusp stars in this region, we computed another model including a stellar cusp extending beyond the outer edge of the discs, which did not yield significantly higher eccentricities.

\begin{figure}
  \begin{center}
    \includegraphics[width=8.3cm]{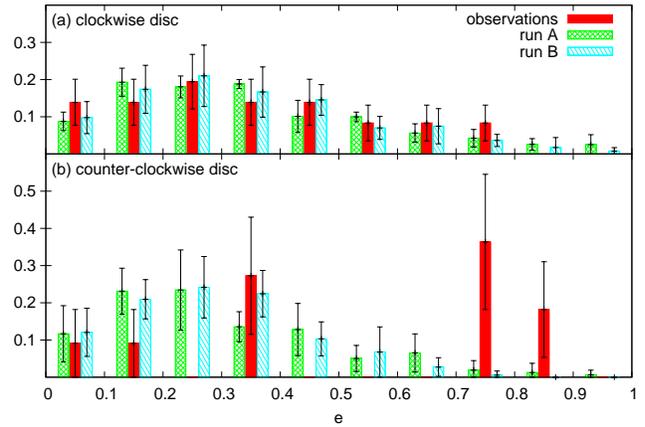}
  \end{center}
  \caption{Eccentricity distribution of stars within 0.3\,pc from the SMBH.
  The green and cyan boxes show the results of our models $A$ and $B$, respectively, each averaged over five different realisations with standard deviation shown as error bars.
  The red filled boxes depict the values derived by \citet{pau06} for the observed discs with Poisson error bars.
  (a) The eccentricity distribution of the observed clockwise disc is in good agreement with our simulations.
  (b) In the observed counter-clockwise system, 6 out of 11 stars have eccentricities $e>0.7$, which is not consistent with our simulations.
    \label{fig:ecc-hist}}
\end{figure}

Fig.\ \ref{fig:ecc-hist} shows the eccentricity distribution of stars within 0.3\,pc from the SMBH.
While the observed eccentricities of the clockwise disc are very well resembled by our simulations (upper panel),
the fraction of 6 out of 11 stars with observed eccentricities $e>0.7$ in the counter-clockwise system is significant and not consistent with any of our two models.
However, integrating a model where the less massive disc starts with the observed eccentricity distribution results in a consistent eccentricity distribution of this disc while retaining its warping and the other disc's properties.
We thus conclude that the stars in the counter-clockwise system probably formed from an eccentric disc, which has been shown by \citet{br08} to be a viable scenario.

\section{Discussion and Outlook}
Our results show that the orbits of young stars observed within 0.3\,pc from Sgr A$^*$ can be very well explained by the interaction of two initially flat discs.
While the present existence of a counter-clockwise disc in the GC is still controversial, the observed structures do not exclude but naturally result from a setup of two inclined accretion discs.
We find that this scenario especially accounts for the controversial nature of the counter-clockwise disc (i.e., a higher disc thickness), as this less massive system is exposed to stronger warping and dissolution.
Especially towards the inner 0.1\,pc, this warping is so strong that one cannot expect to find a disc structure after 5 Myr of interaction.
Hence, the two-disc scenario is consistent with \citet{lu08} not finding a counter-clockwise disc in their recent data set extending to 0.1\,pc.
Whether or not the less massive stellar system is called a disc may be a matter of discussion, even if it initially had a perfect disc shape.

From our simulations we see that both non-resonant as well as (short-term) resonant relaxation effects play a significant role in the evolution of orbital eccentricities of the disc stars, although the exact contributions are not yet fully understood. Furthermore, we see that both disc and cusp stars contribute significantly to relaxation.

We are left with the question of how two mutually highly inclined dense accretion discs can have formed (almost) simultaneously a few Myr ago.
Furthermore, our models do not explain the relatively large eccentricities and inclinations of the orbits towards the outer edge of the stellar disc(s).
For this, one may consider the effects of an external perturbation, or simply argue that the uncertainties of the observations of these stars are comparably high. In addition, \citet{br08} have shown how eccentric stellar discs can be created from high-eccentricity infalls of giant molecular clouds (GMCs).

\citet{hn08} simultaneously explained both the double accretion disc and the diffusion at larger radii: Their calculations show how (self-)colliding GMCs get bound to the SMBH and may form two accretion discs at large inclination in the vicinity of the SMBH. At somewhat larger distances, the geometry is much more complex, which may account for the observed larger eccentricities and inclinations.

We suspect that a thorough analysis of the orbital parameters of young stars in the GC will contribute significantly to our understanding of star formation under such extreme circumstances.
Our calculations rule out the existence of two 5\,Myr old flat discs within the inner 0.3\,pc from Sgr A$^*$. None the less, detailed properties of the stellar orbits as obtained e.g.\ by analyses of stellar accelerations may help to describe the disc warping and eccentricity distribution more precisely and thus shed light on the exact gas infall scenario. Furthermore, the detection of B-type stars in the discs is expected to increase the significance of the disc warping (and of the counter-clockwise disc, as larger data sets may yield the opportunity to confirm it despite massive warping by constraining to subsets of smaller radial extent) and yield more information about the discs' stellar mass function and total mass.

\section*{Acknowledgments}
We thank the anonymous referee for helpful comments.
This work was supported by the German Research Foundation (DFG) through the priority program 1177
`Witnesses of Cosmic History: Formation and Evolution of Black Holes, Galaxies and Their Environment'.

\makeatletter   \renewcommand{\@biblabel}[1]{[#1]}   \makeatother

\label{lastpage}

\end{document}